\documentclass[nofootinbib,longbibliography,floatfix,superscriptaddress,twocolumn]{revtex4-1}

\usepackage[english]{babel}
\usepackage{amsmath,amssymb,amsbsy,amstext,amsthm,booktabs,simplewick,exscale,relsize,slashed,graphicx,amsfonts,upgreek}
\usepackage{tabu,multirow,color,dcolumn,bm,enumerate}
\usepackage{tikz}
\usetikzlibrary{decorations.pathmorphing,,decorations.markings}

\begin{document}

\title{Dark matter ignition of type Ia supernovae}

\author{Joseph Bramante} 
\address{Department of Physics, University of Notre Dame, 225 Nieuwland Hall, Notre Dame, IN, USA}


\begin{abstract}
Recent studies of low redshift type Ia supernovae (SNIa) indicate that half explode from less than Chandrasekhar mass white dwarfs, implying ignition must proceed from something besides the canonical criticality of Chandrasekhar mass SNIa progenitors. We show that $1-100$ PeV mass asymmetric dark matter, with imminently detectable nucleon scattering interactions, can accumulate to the point of self-gravitation in a white dwarf and collapse, shedding gravitational potential energy by scattering off nuclei, thereby heating the white dwarf and igniting the flame front that precedes SNIa. We combine data on SNIa masses with data on the ages of SNIa-adjacent stars. This combination reveals a $
 2.8 \sigma$ inverse correlation between SNIa masses and ignition ages, which could result from increased capture of dark matter in 1.4 versus 1.1 solar mass white dwarfs. Future studies of SNIa in galactic centers will provide additional tests of dark-matter-induced type Ia ignition. Remarkably, both bosonic and fermionic SNIa-igniting dark matter also resolve the missing pulsar problem by forming black holes in $\gtrsim 10$ Myr old pulsars at the center of the Milky Way.
\end{abstract}

\maketitle

It would be difficult to overstate the significance of type Ia supernovae, which have been used to measure the vacuum energy of the universe. The uniform width-to-height ratio of SNIa emission curves, known as the Phillips relation, allows SNIa distance and redshift to be inferred from luminosity over $\sim$40 day outbursts \cite{Phillips:1993ng}. A prevalent lore states that the uniformity of SNIa light curve width-to-height ratios results from a uniform population of SNIa progenitors: carbon-oxygen white dwarfs (WDs) explode after reaching Chandraskehar mass ($M_{CH} \sim 1.4 ~M_\odot$) by accretion from a binary companion.  

However, a recent multiband study of 20 ``Ia-norm" supernovae light curves, selected from 147 low redshift specimens, presents strong evidence that many type Ia supernovae do not reach $M_{CH}$ before exploding \cite{Scalzo:2014sap}. Follow-on studies have concluded that about half of the observed SNIa have sub-$M_{CH}$ progenitors \cite{Scalzo:2014wxa}, resulting in a relation for SNIa progenitor masses,
\begin{align}
M_{SNIa}/M_\odot = 1.322 \pm 0.022 + (0.185\pm 0.018)x_1, 
\label{eq:stretchmass}
\end{align} 
where $x_1$ parameterizes SNIa light curve stretch and ranges from $-2.8$ to $2.6$. Clearly, these findings are at odds with $M_{CH}$-attainment as the sole mechanism for triggering SNIa. 

There are proposed mechanisms for triggering sub-Chandrasekhar SNIa with known particles, e.g. accreted helium shells can cause ``double-detonations" \cite{Nomoto:1982zz,Yoon:2004zq,Fink:2007fv} or binary WD mergers \cite{Pakmor:2013wia}. These require binary companions, and lone WD SNIa progenitors are preferred by some observations, including the absence of expected luminosity shocks from binary companions \cite{NatureIaCompanion}, little circumstellar material in progenitor systems (corroborated by low initial x-ray and radio emission), and the high overall rate of SNIa \cite{Maoz:2013hna}.

In this paper we demonstrate that heavy asymmetric dark matter (DM) ignites lone WDs by rapidly increasing the WD temperature inside the region of DM collapse. This occurs for halo density DM ($\rho_X \sim \rm{GeV/cm^3}$) collected into a WD within its lifetime. We correlate SNIa masses with nearby stellar ages and find a $2.8 \sigma$ preference for heavier (1.4 $M_\odot$) WDs exploding sooner, which is one prediction of SNIa-igniting DM. 

In the analysis of DM-induced WD ignition that follows, it is sufficient to assume any model with some amount of DM asymmetry (we consider totally asymmetric DM, but partly asymmetric DM only changes the effective capture rate) and a mass around a PeV. These requirements allow enough DM to accumulate, self-gravitate, and collapse within a WD lifetime. We also assume a velocity-independent DM-nucleon cross-section, $\sigma_{nX}$. Note that, because the DM momentum transfer to standard model (SM) particles is less than a GeV throughout, this simplified framework can be UV completed with the addition of weak scale mediators. PeV mass particles can fill out the DM relic abundance (despite unitarity bounds \cite{Griest:1989wd}) through, for example, non-thermal processes \cite{Hall:2009bx} or co-annihilation enhanced freeze-out \cite{Griest:1990kh}.

{\bf 1. DM accumulation.} DM's collection rate in a WD is \cite{Press:1985ug,Gould:1987ju,Goldman:1989nd,Kouvaris:2007ay,Bertone:2007ae}
\begin{align}
C_X =  \frac{\sqrt{24 \pi} G \rho_X  M_{w} R_{w} }{m_X \bar{v}} { \rm Min}\left( 1,\frac{\sigma_{aX}}{\sigma_{sat}} \right)\left[1- \frac{1-e^{-B^2}}{B^2}\right],
\label{eq:cx}
\end{align}
where $G$ is Newton's constant, $\rho_X$ is the DM halo density, $M_w$ and $R_w$ are the WD mass and radius, $m_X$ is the DM mass, $\bar{v} \sim 200~{\rm km/s}$ is the WD-DM velocity dispersion, $\sigma_{sat} \sim R_w^2/N_N$ is the maximum DM-nuclear cross-section ($N_N$ is the number of nuclei in the WD). We take $\hbar=c=k_B=1$ throughout.
The square-bracketed term accounts for DM that scatters but is not captured in the WD, and $B^2= 6 m_X v_{esc}^2/ m_N\bar{v}^2(m_X/m_N-1)^2$, where $v_{esc} \simeq \sqrt{2 G M_w/R_w}$ is the escape velocity from the WD surface and $m_N \simeq 14~\rm{GeV}$ is the average WD nuclear mass. The next sections discuss the coherent scattering form factor incorporated in the DM-nucleus cross-section, $\sigma_{aX}$. (If the DM-WD momentum transfer exceeds the inverse nuclear radius $(1/3~\rm{fm}^{-1})$ during capture, we calculate DM capture on nucleons instead of nuclei.)

\begin{figure}[h!]
\includegraphics[scale=1.2]{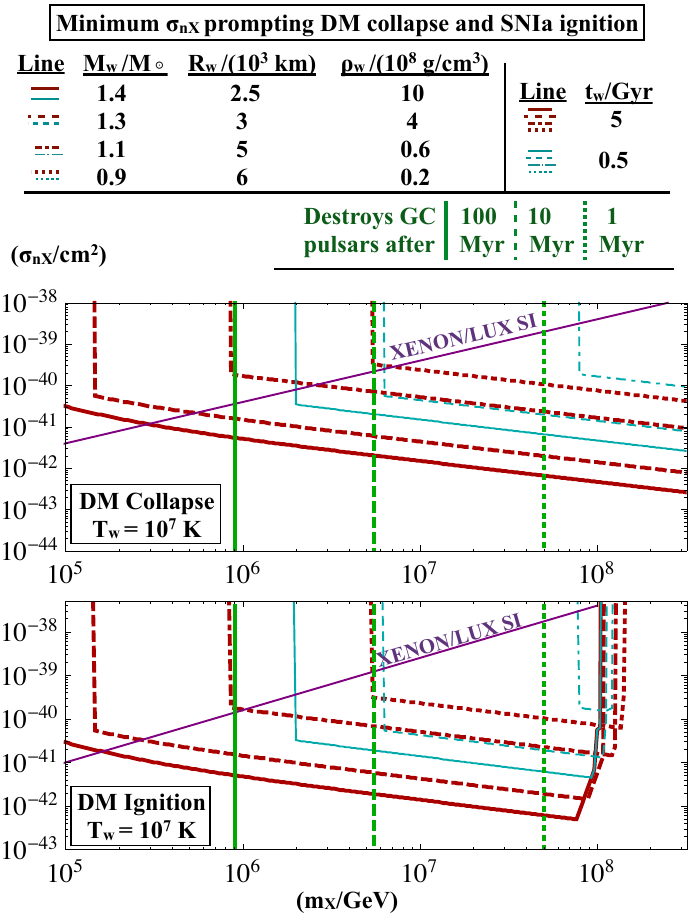}
\caption{The top panel displays the minimum DM-nucleon cross-section necessary for a WD to accumulate a collapsing sphere of DM in $t_w=0.5$ Gyr (thin cyan lines) and $t_w=5$ Gyr (thick red lines). The galactic DM density assumed to surround WDs is $\rho_X \sim \rm{GeV/cm^3}$. Solid, dashed, dotted-dashed and dotted lines correspond to WD masses, radii, and core WD densities indicated.  The bottom panel shows the minimum DM-nucleon cross-section required for asymmetric DM to ignite SNIa after collecting and collapsing in WD progenitors with central temperatures of $T_w = 10^7~ {\rm K}$. For lighter DM, the number of DM particles required for collapse can be too large -- the collection rate is limited by the WDs geometric cross-section -- leading to the low mass cutoffs shown. For heavier DM, the number of DM particles required for collapse decreases, along with the rate of energy injected into the WD, leading to a high mass cutoff for WD ignition. These calculations assume that the DM-DM scattering cross-section equals the DM-nucleon cross-section $\sigma_{XX} \sim \sigma_{nX}$, that is we assume the mediator responsible for DM-nucleon scattering induces comparable DM-DM scattering. Each panel also indicates that heavy asymmetric fermionic (or bosonic with quartic $\lambda \sim 1$) DM collapses $\lesssim$Gyr old pulsars in the Milky Way's galactic center, assuming a central parsec DM density $\rho_X^{GC} \sim 10^4 ~{\rm GeV/cm^3}$. The LUX spin-independent $2 \sigma$ bound on DM-nucleon scattering is given in purple \cite{Aprile:2012nq,Akerib:2013tjd}.}
\label{fig:asymbspace}
\end{figure}

After the DM is captured, it will continue to scatter with the WD until it thermalizes at its center. DM with a mass lighter than 100 PeV and cross section $\sigma_{nX} \gtrsim 10^{-42} ~\rm{cm^2}$ will thermalize in less than a hundred million years \cite{Kouvaris:2010jy}. The thermalized DM sphere will have a radius given by the Virial theorem,
\begin{align}
r_{th} &= (9 T_w / 4 \pi G \rho_w m_X )^{1/2} \nonumber \\ &\simeq 90 ~{\rm m}~ \left(\frac{m_X}{\rm{PeV}}\right)^{-1/2}\left(\rho_{w8}\right)^{-1/2} \left(\frac{T_w}{\rm{10^7~K}}\right)^{1/2},
\label{eq:rtherm}
\end{align} 
where $\rho_{w8} \equiv \rho_w / (10^8~\rm{g/cm^3})$, and $\rho_w = 10^{7}-10^9 ~\rm{g/cm^3}$ is the density of a sub-Chandrasekhar, ignitable WD. If the DM particles are heavier, the thermal radius will be smaller, and fewer DM particles need collect within a WD lifetime to become self-gravitating and collapse. The amount of DM required for collapse is 
\begin{align}
N_{sg} &= 4 \pi \rho_w r_{th}^3/3m_X \simeq 2 \times 10^{38} \nonumber \\ & \times \left(\frac{m_X}{\rm{PeV}}\right)^{-5/2} \left(\rho_{w8}\right)^{-1/2} \left( \frac{T_w}{10^7~\rm{K}}\right)^{3/2} .
\label{eq:nsg}
\end{align} In Figure \ref{fig:asymbspace} we show the DM-nucleon cross-section required for DM to collapse inside $M_w = 0.9-1.4 ~M_\odot$ WDs within $0.5$ and $5$ gigayears.

{\bf 2. DM collapse.} As the sphere of DM collapses, it will shed gravitational potential energy by scattering off carbon and oxygen nuclei in the WD, which can prompt SNIa. The rate and mode of DM collapse is determined by whichever of the gravitational free-fall time and the DM-DM interaction time is shorter \cite{Goldman:1989nd}. At the onset of collapse in a WD, weakly-interacting DM will not be self-thermalized. As collapse progresses, the dynamical free-fall time (which scales with the DM sphere's radius as $\propto r^{3/2}$) will come to exceed the DM-DM scattering time ($\propto r^{7/2}$). Once the free-fall time exceeds the DM-DM interaction time, the DM will self-thermalize through DM-DM interactions, and (in the absence of a more efficient radiation mechanism) shed gravitational energy by nuclear scattering as it collapses further. We will see that self-thermalized, collapsing DM heats the WD enough to spark SNIa.

In total, there are three relevant timescales for the collapsing DM sphere: the dynamical free-fall time, the DM-DM interaction time, and the DM-nuclear interaction time. The free-fall time is
\begin{align}
t_{ff} \sim \sqrt{3 \pi/(32 G \rho_X)}  \simeq 0.15~{\rm s} ~(\rho_{w8})^{-1/2} \left( \frac{r}{r_{th}} \right)^{3/2}.
\label{eq:tff}
\end{align}
Note that at the time of collapse, the DM density equals the WD density. 

The DM-DM interaction time is
\begin{align}
t_{XX} &\sim (n_X \sigma_{XX} v_X)^{-1} \simeq 3 \times  10^{9}~{\rm s}~\left(\frac{m_X}{{\rm PeV}}\right)^{3/2} (\rho_{w8})^{-1} \nonumber \\ &\times \left( \frac{T_w}{10^7~\rm{K}} \right)^{-1/2} \left(\frac{\sigma_{XX}}{{10^{-40} ~\rm{cm^2}} }\right)^{-1}  \left(\frac{r}{r_{th}} \right)^{7/2},
\label{eq:txx}
\end{align}
where $n_X$ is the DM number density and $v_X = \sqrt{2G N_{sg} m_X/r}$ is the DM velocity for a self-gravitating DM sphere of radius $r$. 

The DM-nucleus interaction time is
\begin{align}
t_{aX} &\sim (n_N \sigma_{aX} v_X)^{-1} \simeq 10^{2}~\rm{s}~  \left(\frac{m_X}{{\rm PeV}}\right)^{1/2} (\rho_{w8})^{-1}  \nonumber \\ &\times\left( \frac{T_w}{10^7~\rm{K}} \right)^{-1/2} \left(\frac{\sigma_{nX}}{{10^{-40} ~\rm{cm^2}} }\right)^{-1} \left(\frac{r}{r_{th}} \right)^{1/2},
\label{eq:tnx}
\end{align}
where the final expression includes the DM-nuclear form factor (at maximum coherence), which we discuss shortly.
Comparing the dependence on the collapsing sphere's radius ($r$) in Eqs. \eqref{eq:tff}, \eqref{eq:txx}, and \eqref{eq:tnx}, we see that heavy DM in WDs begins collapse in free-fall, but will transition to thermalized collapse once the DM-DM interaction time becomes shorter than $t_{ff}$. Setting $t_{ff} = t_{XX}$, we solve for the radius below which the collapsing DM is thermalized by DM-DM interactions,
\begin{align}
r_{sta} &\sim  \sqrt{N_{sg} \sigma_{XX}} \simeq 0.1 ~{\rm cm}~\left(\frac{m_X}{\rm{PeV}}\right)^{-5/4}  \left(\rho_{w8}\right)^{-1/4} \nonumber \\ &\times \left( \frac{T_w}{10^7~\rm{K}}\right)^{3/4}  \left(\frac{\sigma_{XX}}{{10^{-40} ~\rm{cm^2}} }\right)^{1/2}.
\label{eq:rsta}
\end{align}
Prior to collapsing to $r_{sta}$, the DM sphere must discard the difference in gravitational potential energy between $r_{th}$ and $r_{sta}$ by scattering off WD nuclei. Setting $r=r_{th}$ in Eq. \eqref{eq:tnx}, and noting that a DM particle will reach equilibrium with the WD after $\sim m_X /2m_N$ scatters \cite{Lewin:1995rx}, we estimate the maximum time required for the DM to collapse to radius $r_{sta}$,
\begin{align}
t_{th} &\lesssim  t_{aX} m_X/2 m_N \simeq 0.1~\rm{yrs}~  \left(\frac{m_X}{{\rm PeV}}\right)^{3/2} (\rho_{w8})^{-1}  \nonumber \\ &\times\left( \frac{T_w}{10^7~\rm{K}} \right)^{-1/2} \left(\frac{\sigma_{nX}}{{10^{-40} ~\rm{cm^2}} }\right)^{-1}.
\end{align}
In the analysis that follows, we assume $\sigma_{nX} \sim \sigma_{XX}$, that is, that the same mediator responsible for DM-nucleon scattering also mediates DM-DM scattering, giving both processes about the same cross-section. Relaxing this assumption, particularly $\sigma_{XX} \ll \sigma_{nX}$, results in other DM-igniting SNIa parameter space, as we will discuss. 

{\bf 3. SNIa ignition.} Once the DM has self-thermalizes (at $r_{sta}$), it uniformly heats the WD (within $r_{sta}$) as it collapses further. First note that the DM-WD momentum transfer is smaller than the inverse of the WD nuclear radii, $p \sim m_N v_{sta} < (3 ~{\rm fm})^{-1}$ where $v_{sta} = \sqrt{2G N_{sg} m_X/r_{sta}}$. (It can also be verified that $v_{sta}$ is the relative velocity of the DM-WD nuclear system.) This indicates that scattering off WD nuclei will be coherently enhanced. The DM-nucleus cross-section is $\sigma_{aX} \simeq A^2 (3j_1[x]/x)^2 {\rm Exp}(-x^2/3) \sigma_{nX}$, where $A^2=200$ accounts for coherent scattering enhancement off carbon and oxygen nuclei, $j_1$ is the Bessel function of the first kind, $x \equiv p r_n$, and $r_n \simeq 3~\rm{fm}$ \cite{Primack:1988zm,Lewin:1995rx}. The average energy transferred per DM-nuclei scatter is then $\epsilon \sim m_N v_{sta}^2/2$, occurring on a timescale set by $t_{sta} = t_{aX}(r_{sta})$. The rate of energy transferred to the WD is
\begin{align}
&\dot{Q}_{he} = N_{sg}  \epsilon /t_{sta} \simeq 2 \times 10^{33} ~{\rm GeV/s}\left(\frac{m_X}{{\rm PeV}}\right)^{-23/8} (\rho_{w8})^{1/8}  \nonumber \\ &\times\left( \frac{T_w}{10^7~\rm{K}} \right)^{21/8} \left(\frac{\sigma_{nX}}{{10^{-40} ~\rm{cm^2}} }\right)\left(\frac{\sigma_{XX}}{{10^{-40} ~\rm{cm^2}} }\right)^{-3/4},
\label{eq:qsta}
\end{align}
where this assumes coherent DM-nuclei scatters, but we use the preceding nuclear form factors in computations.

If $\dot{Q}_{he}$ is larger than the rate at which the heat diffuses in the WD, the collapsing DM can ignite SNIa. The work of \cite{Timmes:1992} showed that for WD material of mass $m_{he}$, in the mass range $10^{-5} < (m_{he} / {\rm g}) < 10^{15}$, SNIa ignition requires heating a mass $m_{he}$ of carbon-oxygen to temperature $(T_{he} / 10^{9.7}~{\rm K})^{70/3} \gtrsim (\rho_{w8})^{1/2}({\rm g} / m_{he})$. A close inspection of Eq. \eqref{eq:rsta} reveals that heavier DM in cooler WDs will enclose less than a milligram of WD at $r_{sta}$. Therefore, to remain well within the numerical calculations of \cite{Timmes:1992}, we consider heat diffusion out of a sphere of radius $r_{he}$, where $r_{he}$ is the larger of $r_{sta}$ and a sphere enclosing a milligram of WD, $r_{mg} = 1.5\times 10^{-4} ~\rm{cm}~(\rho_{w8})^{-1/3}$. Then for SNIa ignition, the DM heat transfer (Eq. \eqref{eq:qsta}) must exceed the conductive diffusion rate out of a $r_{he}$ size sphere at the WD center, given by \cite{Shapiro:1983du}
\begin{align}
\dot{Q}_{dif} \simeq 4 \pi^2 r_{he} T_{he}^3 (T_{he} - T_w)/15 \kappa_c \rho_w,
\label{eq:qdif}
\end{align}
where $\kappa_c \sim (10^{-7} ~\rm{cm^2/g} ) (T_{he}/10^7~ \rm{K})^{2.8} (\rho_{w8})^{-1.6}$ is the conductive opacity for WDs, which have a thermal diffusion dominated by relativistic electron conduction when $\rho_w \gtrsim 10^{6}~{\rm g/cm^3}$ \cite{Shapiro:1983du}. The preceding expression for $\kappa_c$ conforms to WD conductive opacity tables over the relevant range of WD temperatures and densities \cite{Potekhin:1999yv,PotekhinUrl}. 

In Figure \ref{fig:asymbspace} we display DM masses and minimum cross-sections required to ignite WDs within 0.5 and 5 Gyr. In addition to $\dot{Q}_{he} >\dot{Q}_{dif} $, we require that the DM impart enough heat to bring the WD core temperature to $T_{he}$. This requirement is easily fulfilled if $\dot{Q}_{he} >\dot{Q}_{dif} $, because the capacitance of a degenerate WD's ion lattice is simple ($c_v = 3k_b/2$ \cite{Shapiro:1983du}), so in the limiting case of slowest heating ($m_X \sim 100~\rm{PeV}$, $\rho_w \simeq  10^9~\rm{g/cm^3}$, and $\sigma_{nX} \simeq 10^{-42} ~\rm{cm^2}$), the DM will heat a gram of WD to $10^{10} ~\rm{K}$ (1 MeV) after a fraction of collapsing DM particles have scattered once, $N_{sca} \sim (10^{23} ~{\rm nuclei/g \rm})({\rm MeV} / \epsilon N_{sg}) \lesssim 10^{-7}$. The curves in Fig. \ref{fig:asymbspace} cut off at high mass, because too few DM particles collapse, c.f. Eq. \eqref{eq:nsg}, to adequately heat the WD.

It can be shown that PeV mass fermionic DM is not degenerate (and is not Pauli-blocked) as it collapses through $r_{sta}$. To calculate the radius at which the DM becomes degenerate, first note that the most energetic stabilized fermions have minimum Fermi kinetic energy $E_f = (9 \pi N_{sg}/4 )^{2/3}/2 m_X r^2$. This implies a DM radius (using the Virial theorem),
\begin{align}
r_{deg} &= (9\pi /4)^{2/3}/G m_X^3 N_{sg}^{1/3} \simeq  10^{-4}~\rm{cm} \nonumber \\
 &\times \left(\frac{T_w}{10^7~\rm{K}}\right)^{-1/2} \left(\frac{m_X}{\rm{PeV}}\right)^{-13/6} \left(\rho_{w8}\right)^{1/6},
\label{eq:rdeg}
\end{align}
smaller than $r_{sta}$ by two orders of magnitude. A similar computation reveals that collapsing, heavy bosonic DM will not condense until $r \ll r_{sta}$. 

On the other hand, if $\sigma_{XX} \ll \sigma_{nX}$ (we note again that this analysis makes the simplifying assumption that $\sigma_{XX} \sim \sigma_{nX}$), the radius at which the DM self-thermalizes will be smaller. In this limit, we have found that fermionic DM collapsing into a degenerate sphere (and bosonic DM condensing into a Bose-Einstein condensate \cite{Kouvaris:2012dz}), will heat and ignite SNIa for much of the parameter space indicated in Figure \ref{fig:asymbspace}.

\begin{figure}
\includegraphics[scale=.65]{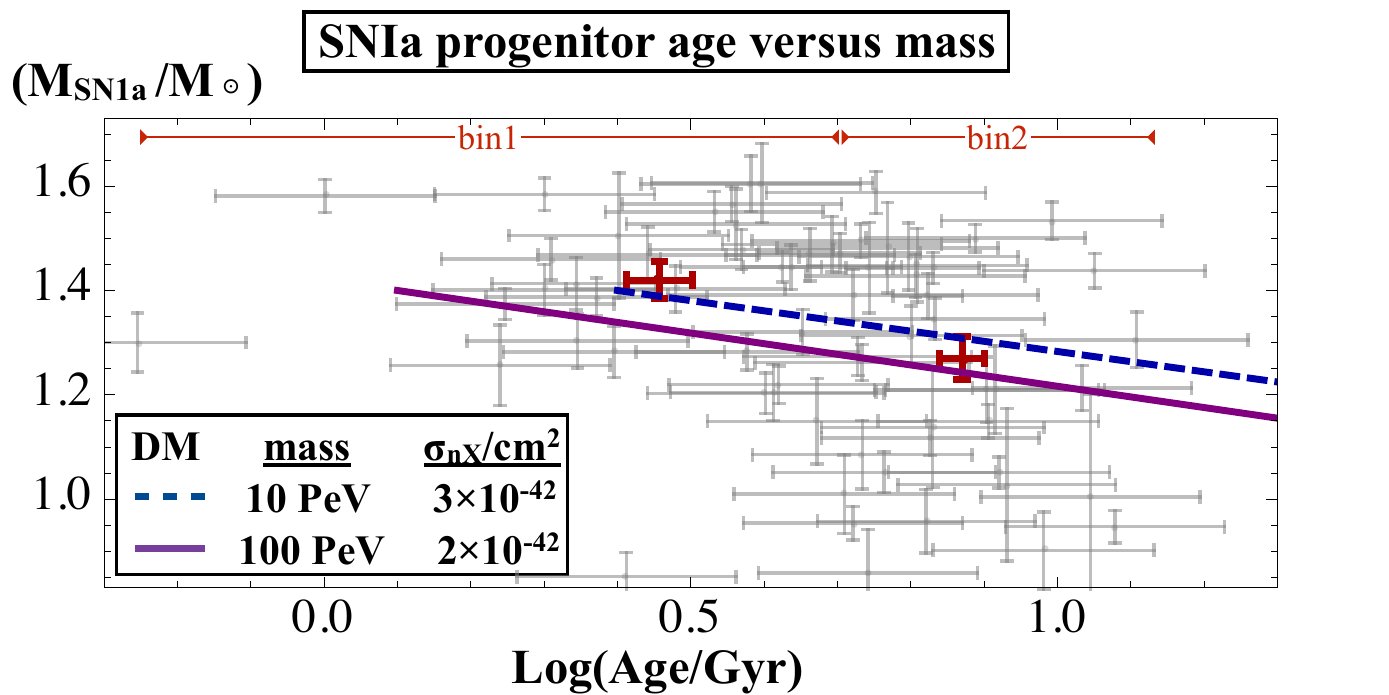}
\caption{Data bins (with $1\sigma$ error bars) are given for SNIa age versus progenitor mass, by combining the SNIa-adjacent star age data in \cite{Pan:2013cva} with the progenitor mass fitting function of \cite{Scalzo:2014wxa}. Asymmetric DM model curves for SNIa-ignition age as a function of progenitor mass are overlaid (these were obtained using the same methods as Figure \ref{fig:asymbspace}). Note that the DM model curves assume uniformly carbon-oxygen WDs with temperature $10^7$ K.}
\label{fig:correlation}
\end{figure}

{\bf 4. SNIa age vs. mass.}
Results in prior sections indicate that if SNIa are triggered by PeV mass DM, this implies a negative correlation between SNIa progenitor age and mass. In Figure \ref{fig:correlation}, we reframe a study that correlated host galaxy star age with light curve stretch \cite{Pan:2013cva} (see also \cite{Childress:2013wna,Childress:2013xna,Rigault:2013gux}), by converting light curve stretch to SNIa mass with results from \cite{Scalzo:2014wxa} (see Eq.~\ref{eq:stretchmass}). Refs.~\cite{Pan:2013cva} and \cite{Scalzo:2014wxa} use SNIa stretch parameters $s$ and $x_1$ respectively; Refs.~\cite{Guy:2007dv,Guy:2010bc} convert $s$ to $x_1$.

In Figure \ref{fig:correlation}, 67 SNIa data points from Ref.~\cite{Pan:2013cva} are collected in bins spanning $0.5-5$ and $5-14$ Gyr. Our analysis of vertical error bar heights agrees with \cite{Pan:2013cva}. The vertical separation between the two mass bins, accounting for uncertainty introduced converting $s$ to $x_1$ to $M_{SNIa}$, is $0.15 ~{\rm M_{\odot}}$ ($M_{bin1} = 1.42 \pm 0.035 ~{\rm M_{\odot}}$, $M_{bin2} = 1.27 \pm 0.041~ {\rm M_{\odot}}$), amounting to a $2.8 \sigma$ significant separation, comparable to the $3.3 \sigma$ result in \cite{Pan:2013cva}, which correlates age with $s$ instead of SNIa mass. We overlay model curves for SNIa-igniting asymmetric DM. 

{\bf 5. Galactic center pulsar implosions.}
The center of the Milky Way does not harbor as many pulsars as expected \cite{Dexter:2013xga,Chennamangalam:2013zja}. Asymmetric DM, more dense in the galactic center (GC), could abundantly collect in GC pulsars and form pulsar-destroying black holes, as first noted in \cite{deLavallaz:2010wp}. DM models fitting the missing pulsar anomaly were presented in \cite{Bramante:2014zca,Bramante:2015dfa}, and this putative population of imploding pulsars could also be the source of fast radio bursts \cite{Fuller:2014rza}. Figure \ref{fig:asymbspace} shows that PeV mass asymmetric DM would destroy 0.1 Gyr old pulsars at the GC, where we use the same pulsar calculations as \cite{Bramante:2014zca,Bramante:2015dfa}. 

The pulsar-imploding parameter space in Figure \ref{fig:asymbspace} is insensitive to $\sigma_{nX}$, because the pulsar DM capture cross-section saturates when $\sigma_{nX}\sim 10^{-45}~\rm{cm^2}$, meaning the amount of DM collected in GC pulsars is constant for $\sigma_{nX} \gtrsim 10^{-45}~\rm{cm^2}$. With the mass of collected DM remaining constant, heavier DM fermions will form black holes in pulsars while lighter DM will not, because the critical mass necessary for black hole formation drops as the DM mass increases, $M_{crit}^{ferm} \sim M_{pl}^3/m_X^2$. PeV mass bosonic DM with an order one quartic self-coupling, $\lambda \sim 1$, will have the same critical mass for forming a black hole as fermionic DM, $M_{crit}^{bos} \simeq \sqrt{\lambda} M_{pl}^3/m_X^2$ \cite{Colpi:1986ye}. This is particularly important when comparing these results to studies that assume asymmetric bosonic DM with a vanishing quartic ($\lambda \lesssim 10^{-15}$)  \cite{McDermott:2011jp,Kouvaris:2011fi,Guver:2012ba,Kouvaris:2012dz,Bramante:2013hn,Bell:2013xk,Bertoni:2013bsa,Zheng:2014fya}. Note also that the cross-sections we consider are smaller than those relevant for other stellar probes of DM (e.g. main-sequence \cite{Taoso:2010tg,Frandsen:2010yj,Iocco:2012wk,Lopes:2012af,Zentner:2011wx,Vincent:2014jia}, astroseismic \cite{Casanellas:2012jp,Casanellas:2015uga,Brandao:2015lra}, cooling \cite{Moskalenko:2007ak,Iocco:2008xb,Yoon:2008km,Taoso:2008kw,McCullough:2010ai}, and pulsar phase \cite{Perez-Garcia:2014dra} constraints).

{\bf 6. Conclusions.} We have introduced a mechanism for igniting SNIa. Heavy asymmetric DM with detectable SM interactions can collapse in and heat 0.9-1.4 solar mass WDs, prompting a thermonuclear runaway. We have pointed out a possible inverse correlation between SNIa progenitor masses and ages, which is predicted if DM ignites SNIa. Studies of SNIa in galactic centers, which harbor a denser bath of DM and would prompt younger and less massive WDs to explode, will provide an additional probe of DM-ignited SNIa.
\linebreak

{\bf Note added.} As this paper was being finalized, Ref. \cite{Graham:2015apa} appeared; it also
considers ways, mostly different, that dark matter can trigger
supernovae.
\section*{Acknowledgements}

I thank the anonymous referees for useful comments. It is a pleasure to thank Robert Lasenby for early collaboration, and also Matthew McCullough for early collaboration, useful comments on the manuscript, and the suggestion to consider collapse in the $\sigma_{nX} \sim \sigma_{XX}$ limit. I thank Surjeet Rajaendran for a stimulating, preliminary discussion regarding local heating of WDs by primordial black holes. I thank James Bramante, Fatemeh Elahi, Roni Harnik, Jason Kumar, Rafael Lang, Adam Martin, and especially Peter Garnavich for useful discussions and correspondence. This research was supported in part by Perimeter Institute for Theoretical Physics. I thank the CERN theory division and University of Heidelberg for hospitality while portions of this work were completed. 

\bibliography{WhiteDwarfDMref}

\end{document}